\documentclass[sigconf]{acmart}
\usepackage{bm}
\usepackage{amsmath}
\usepackage{optidef}
\usepackage{cleveref}
\usepackage{subfigure}
\usepackage{multirow}
\usepackage{graphicx}
\AtBeginDocument{%
  }

\setcopyright{acmlicensed}
\copyrightyear{2024}
\acmYear{2024}
\setcopyright{acmlicensed}
\acmConference[ICMR '24] {Proceedings of the 2024 International Conference on Multimedia Retrieval}{June 10--14, 2024}{Phuket, Thailand.}
\acmBooktitle{Proceedings of the 2024 International Conference on Multimedia Retrieval (ICMR '24), June 10--14, 2024, Phuket, Thailand}
\acmISBN{979-8-4007-0619-6/24/06}
\acmDOI{10.1145/3652583.3657614}
\begin{document}

\title{TIM: Temporal Interaction Model in Notification System}

\author{Huxiao Ji}
\email{jihuxiao@kuaishou.com}
\affiliation{%
  \institution{Kuaishou Technology Co., Ltd.}
  \city{Beijing}
  \country{China}
}
\author{Haitao Yang}
\email{yanghaitao@kuaishou.com}
\author{Linchuan Li}
\email{lilinchuan@kuaishou.com}
\affiliation{%
  \institution{Kuaishou Technology Co., Ltd.}
  \city{Beijing}
  \country{China}
}
\author{Shunyu Zhang}
\email{zhangshunyu@kuaishou.com}
\affiliation{%
  \institution{Kuaishou Technology Co., Ltd.}
  \city{Beijing}
  \country{China}
}
\author{Cunyi Zhang}
\email{zhangcunyi@kuaishou.com}
\affiliation{%
  \institution{Kuaishou Technology Co., Ltd.}
  \city{Beijing}
  \country{China}
}
\author{Xuanping Li}
\authornote{corresponding author}
\email{lixuanping@kuaishou.com}
\affiliation{%
  \institution{Kuaishou Technology Co., Ltd.}
  \city{Beijing}
  \country{China}
}
\author{Wenwu Ou}
\email{ouwenwu@gmail.com}
\affiliation{%
  \institution{Independent}
  \city{Beijing}
  \country{China}
}

\renewcommand{\shortauthors}{Huxiao Ji et al.}


\begin{abstract}
    Modern mobile applications heavily rely on the notification system to acquire daily active users and enhance user engagement. Being able to proactively reach users, the system has to decide when to send notifications to users. Although many researchers have studied optimizing the timing of sending notifications, they only utilized users' contextual features, without modeling users' behavior patterns. Additionally, these efforts only focus on individual notifications, and there is a lack of studies on optimizing the holistic timing of multiple notifications within a period. To bridge these gaps, we propose the Temporal Interaction Model (TIM), which models users' behavior patterns by estimating CTR in every time slot over a day in our short video application Kuaishou. TIM leverages long-term user historical interaction sequence features such as notification receipts, clicks, watch time and effective views, and employs a temporal attention unit (TAU) to extract user behavior patterns. Moreover, we provide an elegant strategy of holistic notifications send time control to improve user engagement while minimizing disruption. We evaluate the effectiveness of TIM through offline experiments and online A/B tests. The results indicate that TIM is a reliable tool for forecasting user behavior, leading to a remarkable enhancement in user engagement without causing undue disturbance.
\end{abstract}




\maketitle

\section{Introduction}
In an era where the digital landscape is saturated with an overwhelming amount of content, capturing and maintaining user attention has become a formidable challenge for mobile applications. Amidst the cacophony of digital interactions, mobile phone notifications have emerged as a potent tool for engaging users, acting as a bridge, connecting users to applications and services in a timely, personalized, and contextually relevant manner. However, users typically receive an average of around 100 notifications from different mobile phone applications (apps) per day\cite{mehrotra2015designing}. If the delivery of notifications is poorly timed, users may miss these alerts on their phones. In contrast, if multiple notifications are sent continuously while users are using their phones, they may feel disturbed and consequently disable the notification function. Thus, it is important to find the opportune moments in order to signalize notifications to remind users of important values and increase engagement without causing excessive disturbance. This paper seeks to explore the efficacy of phone notifications as a mechanism to enhance user engagement, delving into the strategies that can optimize their impact without creating undue interference.

\begin{figure*}[htbp]
    \centering
    \includegraphics[width=\linewidth]{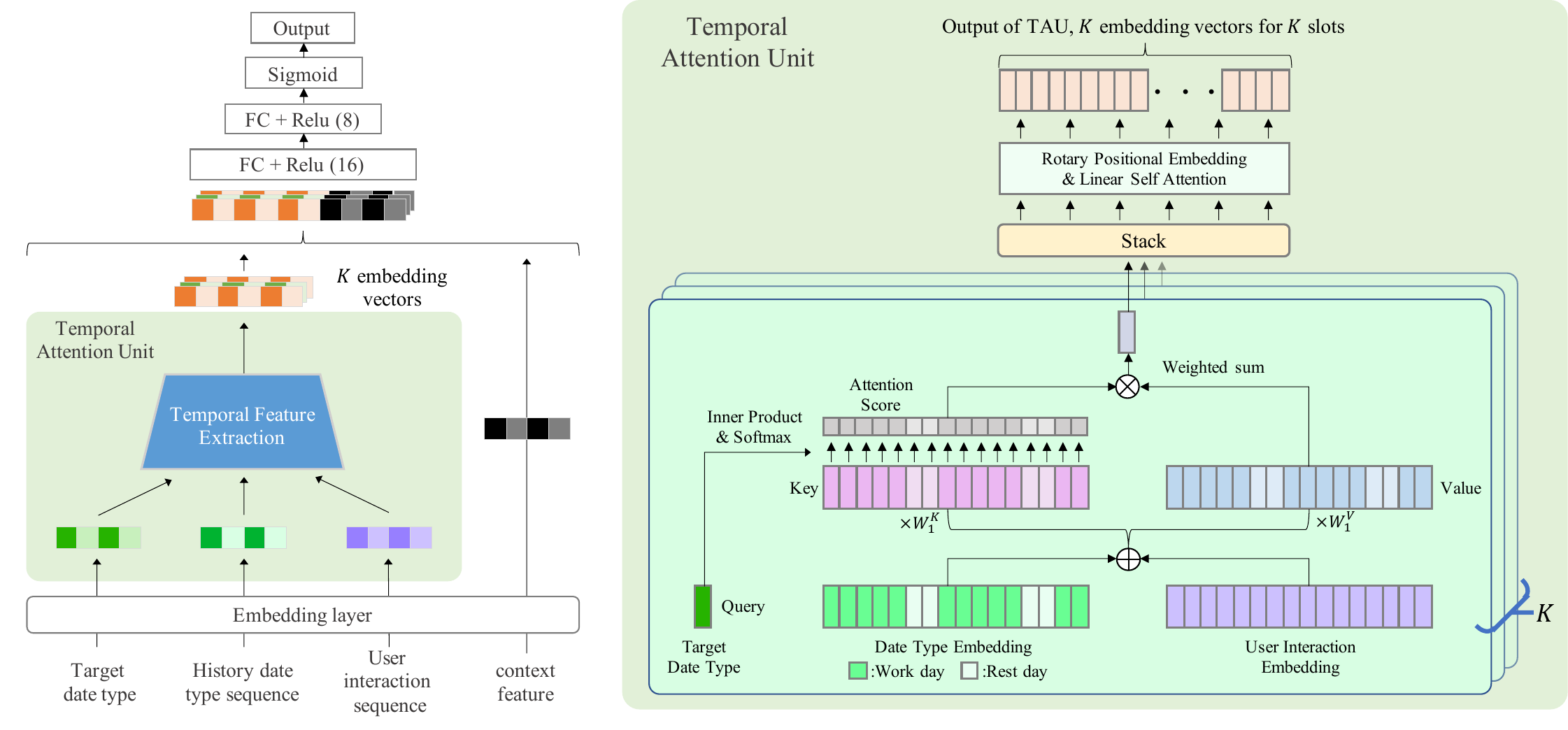}
    \caption{Overview of TIM, with TAU as the central structure for extracting user behavioral patterns. TAU extracts the feature of the target date from the day dimension and then fuses the features in the slot dimension.}
    \Description{Model overview.}\label{pic:model}
\end{figure*}

Recently, a multitude of studies have explored various dimensions of forecasting user engagement on social platforms\cite{gupta2016email,yuan2019state,yuan2022offline,prabhakar2022multi,yang2023deeppvisit}. These methods include analyzing user behavior and social attributes\cite{yang2018know}, constructing user action graphs\cite{liu2019characterizing}, identifying the periodic nature of user activities\cite{chowdhury2021ceam}, assessing the causal impacts of social influence\cite{zhang2022counterfactual}, and capturing similarities from user interactions\cite{tang2020knowing}. These studies employ a uniform model with static parameters across all users, which may not adequately capture the diverse patterns of user engagement. One approach \cite{de2018new} utilizes decision trees to categorize users into distinct clusters, followed by the application of individual logistic regression models for each cluster to predict user attrition. Another study \cite{yang2017personalized} utilizes matrix factorization techniques to forecast personalized engagement in mobile video streaming. However, these approaches fall short in our context, where extensive user interaction sequence data is present.

Previous studies have suggested numerous techniques to determine the optimal moment to send a notification, largely by predicting its response\cite{o2022should,li2023digmn,pielot2017beyond}. One study employed built-in sensors to detect social context and identified ``breakpoint'', an opportune moment to deliver smartphone notifications\cite{park2017don}. Another investigation found that such a breakpoint occurs when a user pauses their work and has spare time to interact with a significant notification\cite{kamal2021hybrid}. However, user engagement may not be attributed to a single notification, but a sequence of notifications, presenting an attribution challenge for modeling. To our best knowledge, no previous work has resolved the issue of determining the holistic timing for sending multiple notifications to enhance user engagement throughout the day.

To fully model the engagement habits of users, we propose the Temporal Interaction Model (TIM). TIM leverages long-term user historical interaction sequences and employs a temporal attention unit (TAU) to extract key behavioral features, with a particular focus on identifying distinct patterns that emerge during work days and rest days. With a day divided into several equal time slots, TIM estimates users' slot-wise CTR over a day. Moreover, we put forward the first strategy that optimizes the holistic timing of multiple notifications in a day, with the ultimate goal of maximizing the notification click volume of Kuaishou.

In the subsequent sections, we will present theoretical derivation and empirical evidence to support our proposed method in \cref*{sec:model}. \Cref*{sec:exp} introduces the details and results of our offline experiments and online A/B tests on KUAISHOU. Finally, \cref*{sec:con} concludes this work and discusses future works.
\section{Methods}
\begin{table*}[t]
    \caption{Performance comparison between TIM and baseline models. Each row represents xgBoost without and with user interaction features, MLP without and with user interaction features and TIM, respectively.}
    \begin{tabular}{l|cccccccccccccc}
        \hline
        \hline
        \multirow{2}{*}{Method} & \multicolumn{7}{c}{Work day} & \multicolumn{7}{c}{Rest day}                                                                                                                                                                                                                                    \\
                                & AUC                         & HR@1                        & HR@5           & HR@9           & A@1            & A@5            & \multicolumn{1}{c|}{A@9}            & AUC             & HR@1            & HR@5           & HR@9           & A@1            & A@5            & A@9            \\ \hline
        XGB                     & 0.5907                      & 0.0201                      & 0.113          & 0.220          & 0.181          & 0.204          & \multicolumn{1}{c|}{0.221}          & 0.5917          & 0.0196          & 0.109          & 0.216          & 0.188          & 0.208          & 0.229          \\
        $\text{XGB}_{int}$      & 0.6673                      & 0.0207                      & 0.118          & 0.227          & 0.187          & 0.212          & \multicolumn{1}{c|}{0.227}          & 0.6644          & 0.0201          & 0.114          & 0.224          & 0.192          & 0.218          & 0.238          \\
        MLP                     & 0.7704                      & 0.0318                      & 0.159          & 0.312          & 0.287          & 0.286          & \multicolumn{1}{c|}{0.281}          & 0.7723          & 0.0310          & 0.154          & 0.306          & 0.295          & 0.294          & 0.291          \\
        $\text{MLP}_{int}$      & 0.8485                      & 0.0391                      & 0.183          & 0.347          & 0.352          & 0.330          & \multicolumn{1}{c|}{0.313}          & 0.8398          & 0.0369          & 0.174          & 0.334          & 0.351          & 0.332          & 0.318          \\
        TIM                     & \textbf{0.8614}             & \textbf{0.0403}             & \textbf{0.188} & \textbf{0.356} & \textbf{0.363} & \textbf{0.339} & \multicolumn{1}{c|}{\textbf{0.320}} & \textbf{0.8559} & \textbf{0.0379} & \textbf{0.179} & \textbf{0.341} & \textbf{0.360} & \textbf{0.340} & \textbf{0.325}
        \\ \hline\hline
    \end{tabular}
    \label{tab:offline}
\end{table*}

\label{sec:model}
In this section, we provide a detailed description of TIM in \cref{sec:mod} and present the derivation and implementation of our strategy in \cref{sec:str}.
\subsection{Temporal Interaction Model}
\label{sec:mod}
To simplify the problem of notification send timing, we divide a day into $K$ equal time slots with length $T$ each and use TIM to estimate users' slot-wise CTR. The overview of TIM is shown in \cref*{pic:model}. We collected various interaction features of users during $K$ time slots in the past $L$ days, including notification receipts, clicks, active duration, video browsing counts, etc. The feature vectors forms a $K\times L\times d_1$ tensor $X=(X_1,X_2,\ldots,X_K)$. We leverage the effectiveness of attention mechanisms\cite{bahdanau2014neural,vaswani2017attention,dosovitskiy2020image,devlin2018bert} for sequence processing by introducing the Temporal Attention Unit (TAU) to extract user behavior pattern embeddings from long-term historical interaction sequences, which addresses the previous limitation of not fully utilizing users' historical behaviors. The fully connected network that follows integrates user contextual features and behavior pattern embeddings.

TAU consists of two attention layers. The first layer is a target attention layer, which outputs one embedding for each of the $K$ slots by extracting user interaction features of the same time slot in previous $L$ days. The outputs are formulated as
$$\bm{o}^1_k=softmax(\frac{\bm{q}_1K_1^\top}{\sqrt{d_2}})V_1$$
where $\bm{q}_1 = \bm{d}^\top_qW^Q_1$, $K_1=(D_h+X_k)W^K_1$, $V_1=(D_h+X_k)W^V_1$, Supposing the projection matrices of query, key and value are $W_1^Q,W_1^K$ and $W_1^V$ correspondingly, $\bm{d}_q$ and $D_h$ represent target date type embedding and historical date type embedding (DTE) respectively, which can alleviate the difference in user behavior habits between work days and rest days, and $d_2$ denotes the projection matrices dimension to scale the weight.

The second attention layer, in which $\bm{o}^1_k(k\in\{1,2,\ldots,K\})$ fuses with each other, is a linear self-attention layer\cite{linearatt} with rotary position embedding (RoPE)\cite{su2024roformer}. Let $\mathcal{O}_1 = (\bm{o}_1^1,\bm{o}_2^1,\ldots,\bm{o}_K^1)^\top$ be the input matrix, $W_2^Q,W_2^K$ and $W_2^V$ be projection matrices of this layer, we have $Q_2 = \mathcal{O}_1W_2^Q,K_2 = \mathcal{O}_1W_2^K,V_2 = \mathcal{O}_1W_2^V,$ and the final output of TAU is calculated as:
$$\mathcal{O}_2=\left(\mathcal{J}+\mathcal{G}(Q_2)\mathcal{G}(K_2)^\top\right)V_2$$
where $\mathcal{J}$ represents a matrix of all 1's, $\mathcal{G}(\cdot)$ is the operation that first adds RoPE to query/key matrix and then normalizes each row of the resultant matrix by $\ell_2$. The output $\mathcal{O}_2 = (\bm{o}_1^2,\bm{o}_2^2,\ldots,\bm{o}_K^2)^\top$ indicates the interaction embeddings of every slot extracted by TAU.
Finally, for each slot $k$, $\bm{o}_k^2$ is concatenated with users' contextual embeddings, and fed into the fully connected network to output the predicted CTR $p_k$.

\subsection{Maximizing Click Count With TIM}
\label{sec:str}
With the predicted slot-wise CTR of TIM, it remains challenging to coordinate these results in practical implementation. The majority of prior research only focused on the timing of a single notification in a time window, while in the real-world scenario of Kuaishou, a user is allocated a given quota of notifications per day. Therefore, we need to perform a holistic optimization rather than optimizing individual points. To address this issue, we model the relationship between slot-wise CTR and our business goals, solve for a sending control strategy that maximizes revenue, and provide an approximate implementation approach.

To achieve our business goal, we need to maximize user's overall click count in one day, which is formulated as
\begin{mini}|l|
    {\bm{n}}{-\bm{p}^\top\bm{n} + \lambda\Vert \bm{n}\Vert_2^2}{}{}
    \addConstraint{\Vert\bm{n}\Vert_1}{=q}
    \addConstraint{n_i}{\geq 0,\quad}{i=1,2, \ldots,K} \label{form:min}
\end{mini}
where the objective $\bm{n} = (n_1,n_2,\ldots,n_K)$ represents the expected numbers of notifications sent in each slot, $\bm{p}=(p_1,p_2,\ldots,p_K)$ is the vector of predicted slot-wise CTR and $q$ denotes the quota of notifications assigned to the user. $\bm{p}^\top\bm{n}$ is the expected click count in one day, and $\lambda\Vert \bm{n}\Vert_2^2(\lambda>0)$ is the regularization term that constrains the number of notifications in a single slot, which ensures that users are not disturbed by continuous notifications.

The Lagrangian function is
\begin{align}
    \mathcal{L}(\bm{n},\varphi,\bm{\mu})=-\bm{p}^\top\bm{n} + \lambda\Vert \bm{n}\Vert_2^2+\varphi(\Vert\bm{n}\Vert_1-q)-\bm{\mu}^\top\bm{n}
\end{align}
and the Karush-Kuhn-Tucker(KKT) conditions are derived as:
\begin{align}
    2\lambda n_i-p_i+\varphi-\mu_i & =0, i=1,2,\ldots,K    \label{form:1} \\
    \Vert\bm{n}\Vert_1             & =q    \label{form:2}                 \\
    \bm{\mu}^\top\bm{n}            & =0    \label{form:3}                 \\
    n_i                            & \geq0, i=1,2,\ldots,K                \\
    \mu_i                          & \geq0, i=1,2,\ldots,K
\end{align}
Combining \cref{form:1,form:2,form:3}, we obtain that
\begin{align}
    n_i & = \left\{
    \begin{array}{lr}
        \frac{1}{2\lambda}(p_i-\varphi), & \mu_i=0    \\
        0,                               & \mu_i\neq0
    \end{array}
    \right.
\end{align}
Note that while $\bm{\mu}=\bm{0}$ and $\lambda = \frac{\Vert\bm{p}\Vert_1}{2q}$, we have $\Vert\bm{n}\Vert_1 = \frac{\Vert\bm{p}\Vert_1-K\varphi}{2\lambda} = \frac{q\Vert\bm{p}\Vert_1-Kq\varphi}{\Vert\bm{p}\Vert_1} = q$, which implies that $\varphi=0$ and $n_i \propto  p_i$. With such observation, we are inspired to set $n_i = q\cdot\frac{p_i}{\Vert\bm{p}\Vert_1}$.

To implement the above-mentioned result, we propose to control the sending with cumulative CTR. Assuming that slot $s$ begins at $t_{s-1}$ and ends at $t_s$ ($t_s - t_{s-1} = T$), and the CTR is consistent throughout the slot. The expected number of notifications sent before time $t_{s-1} + \Delta_t$ ($0\leq \Delta_t <T$) is
\begin{align}
    \mathcal{P}(t_{s-1} + \Delta_t)=\left(P_h+\frac{\Delta_t}{T}\cdot p_s\right)\cdot \frac{q}{\Vert\bm{p}\Vert_1}
\end{align}
where $P_h=\sum_{i=0}^{s-1} p_i$ is the sum of predicted CTR of slots before $s$.
Therefore, when a trigger occurs at time $t_{s-1} + \Delta_t$, the probability to send a notification is
\begin{align}
    \min\left(\mathcal{P}(t_{s-1} + \Delta_t) - \mathcal{H}(t_{s-1} + \Delta_t),1\right)
\end{align}
where $\mathcal{H}(\mathcal{T})$ represents the notification count sent before time $\mathcal{T}$.

\section{Experimental Results}
\label{sec:exp}
In this section, we introduce our experiments. \Cref*{sec:set,sec:met,sec:abl} present the settings, results, and ablation studies of our offline experiments. \Cref*{sec:abt} reports the results of online A/B tests.

\subsection{Experimental Settings}
\label{sec:set}
Our dataset consists of actual data collected from Kuaishou. We randomly select 1 million users' notification data and features during a week (from date $t$ to date $t+6$) as the training dataset and use the data from the same set of users on date $t+8$ as the testing dataset. We sampled two datasets according to whether the date $t+8$ is a work day or a rest day.
We select xgBoost with and without user interaction features (XGB \& $\text{XGB}_{int}$) and multi-layer perception with and without user interaction features (MLP \& $\text{MLP}_{int}$) as baseline models.
To evaluate the efficacy of the models, we compute the area under the receiver operating characteristic curve (AUC), top-$k$ hit ratio (HR@$k$) and top-$k$ Accuracy (A@$k$) on the testing dataset.

\begin{table}[t]
    \footnotesize
    \caption{Impact of RoPE and DTE.}
    \begin{tabular}{cccccccc}
        \hline
        \multicolumn{1}{c|}{Method}              & AUC             & HR@1            & HR@5           & HR@9           & A@1            & A@5            & A@9            \\ \hline
        \multicolumn{8}{l}{Work day}                                                                                                                                      \\ \hline
        \multicolumn{1}{l|}{$\text{TIM}_{base}$} & 0.8535          & 0.0393          & 0.184          & 0.348          & 0.355          & 0.331          & 0.313          \\
        \multicolumn{1}{l|}{$\text{TIM}_{R}$} & 0.8556          & 0.0400          & 0.186          & 0.350          & 0.361          & 0.335          & 0.315          \\
        \multicolumn{1}{l|}{$\text{TIM}_{D}$}  & 0.8532          & 0.0393          & 0.184          & 0.348          & 0.354          & 0.331          & 0.313          \\
        \multicolumn{1}{l|}{$\text{TIM}$}        & \textbf{0.8614} & \textbf{0.0403} & \textbf{0.188} & \textbf{0.356} & \textbf{0.363} & \textbf{0.339} & \textbf{0.320} \\ \hline
        \multicolumn{8}{l}{Rest day}                                                                                                                                      \\ \hline
        \multicolumn{1}{l|}{$\text{TIM}_{base}$} & 0.8457          & 0.0375          & 0.177          & 0.337          & 0.357          & 0.336          & 0.321          \\
        \multicolumn{1}{l|}{$\text{TIM}_{R}$} & 0.8464          & 0.0379          & 0.178          & 0.338          & 0.361          & 0.338          & 0.322          \\
        \multicolumn{1}{l|}{$\text{TIM}_{D}$}  & 0.8464          & 0.0376          & 0.177          & 0.337          & 0.357          & 0.336          & 0.321          \\
        \multicolumn{1}{l|}{$\text{TIM}$}        & \textbf{0.8559} & \textbf{0.0379} & \textbf{0.179} & \textbf{0.341} & \textbf{0.360} & \textbf{0.340} & \textbf{0.325} \\ \hline
    \end{tabular}
    \label{tab:abl}
\end{table}
\begin{figure}[b]
    \centering
    \subfigure[TIM]{
        \label{pic:att1}
        \includegraphics[width=0.43\linewidth]{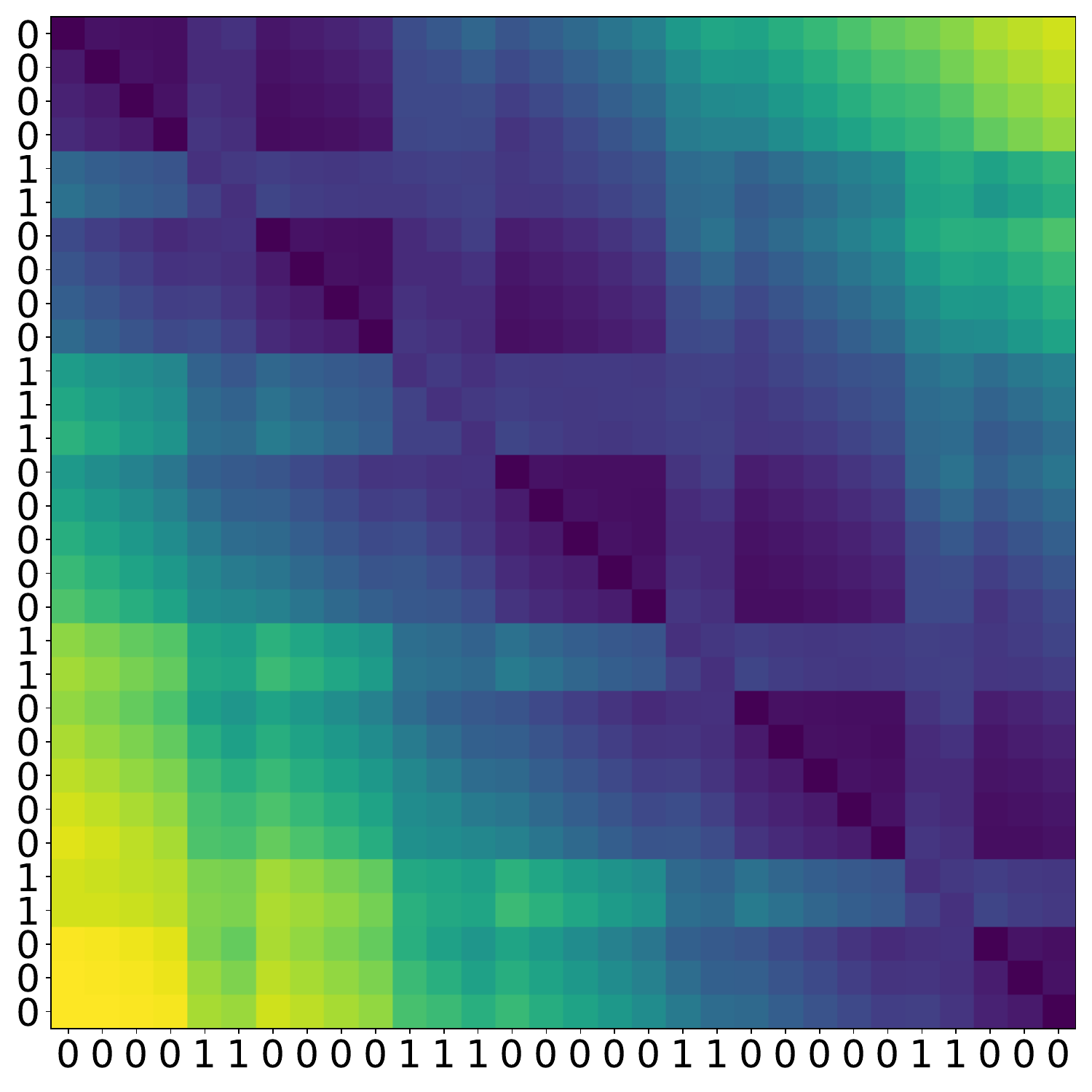}
    }
    \subfigure[$\text{TIM}_{R}$]{
        \label{pic:att2}
        \includegraphics[width=0.43\linewidth]{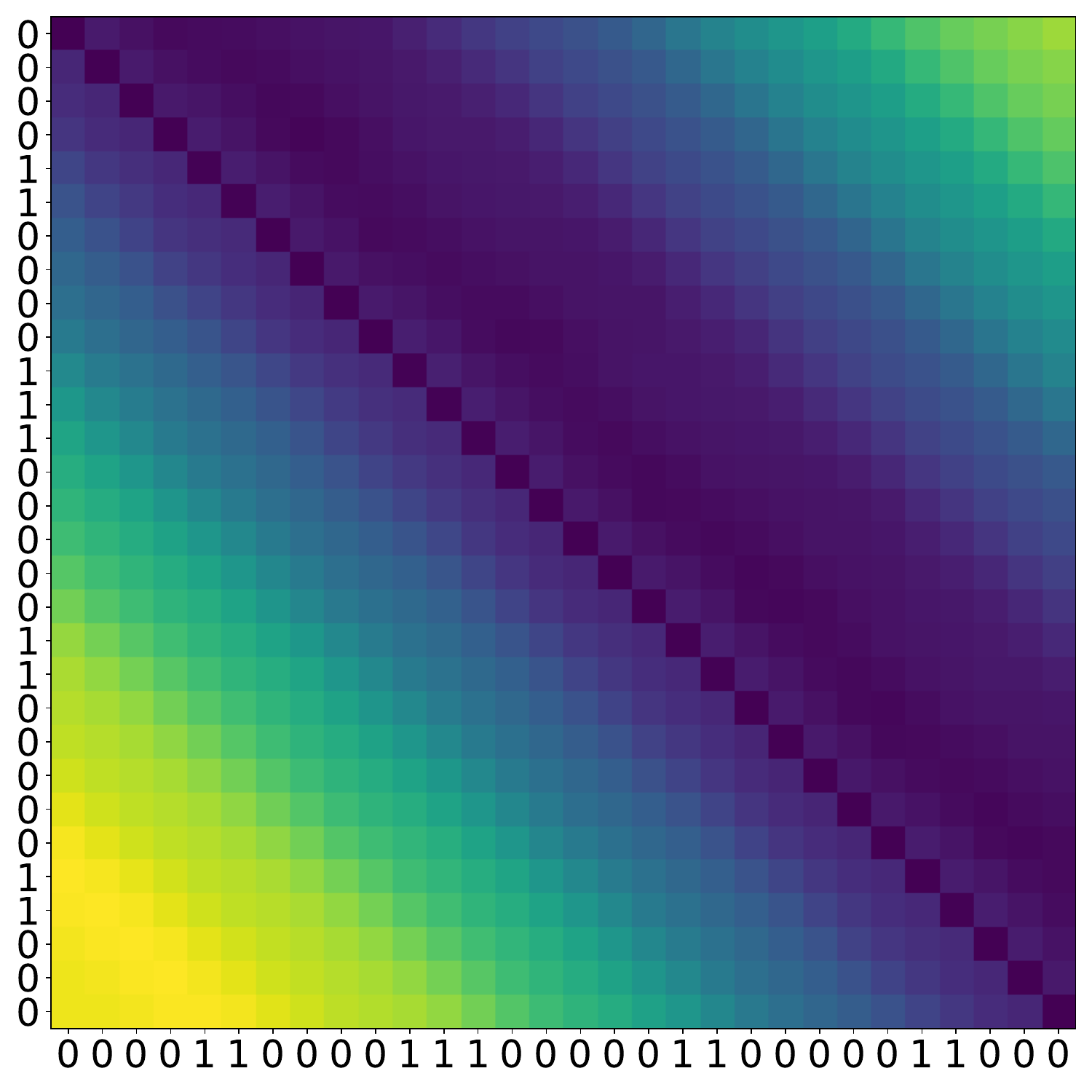}
    }
    \caption{Attention scores in TIM (a) and $\text{TIM}_{R}$ (b).}
    \Description{The attention score heatmap of TIM exhibits clear striped patterns that correlate with work days and rest days, while those of $\text{TIM}_{R}$ do not.}
    \label{pic:att}
\end{figure}
\subsection{Offline Metrics Comparison}
\label{sec:met}
The offline metrics of TIM and baseline models are reported in \cref{tab:offline}. Based on the results, we observe that TIM outperforms all baseline models on all metrics, both on work day and rest day, which shows the strong power of TIM in extracting user behavior patterns. Moreover, $\text{XGB}_{int}$ and $\text{MLP}_{int}$ surpass XGB and MLP respectively by a large margin, indicating the significant effect of user history interaction features.

\subsection{Ablation Study}
\label{sec:abl}
In order to figure out the contribution of RoPE and DTE, we conduct experiments on TIM without RoPE and DTE ($\text{TIM}_{base}$), and with only RoPE ($\text{TIM}_{R}$) or DTE ($\text{TIM}_{D}$). Results are reported in \cref{tab:abl}. We observe the performance of TIM surpasses both $\text{TIM}_{R}$ and $\text{TIM}_{D}$, showing that both RoPE and DTE work in enhancing model ability.

As supporting evidence, we extracted the query embedding for history 30 days and calculated their average attention scores to each other in the target attention layer of TIM and $\text{TIM}_{R}$. The results are shown in \cref{pic:att1,pic:att2} as heat maps. The ``0'' on the axis represents work days and ``1'' represents rest days. We observe that the attention score heatmap of TIM exhibits clear striped patterns that correlate with work days and rest days, while that of $\text{TIM}_{R}$ does not. This suggests that DTE can assist the model in accurately extracting user behavior patterns.
\subsection{Online Evaluation}
\label{sec:abt}
We also report on the online business gains of TIM through the A/B test, where users are randomly divided into three groups. For users in the control group, notifications are distributed evenly over time. Users in another group receive notifications based on the overall slot-wise CTR. The last group uses TIM to control the delivery. For a fair comparison, the quotas of notifications of users in all groups are assigned by the same algorithm. We mainly focus on the following metrics:
\begin{itemize}
    \item{DAU}: Daily active users.
    \item{Watch time}: Total duration spent on watching videos.
    \item{Send volume}: Total notification sent volume.
    \item{CTR}: Click-through rate, the percentage of notifications that get clicked over all notifications in one day.
    \item{Switch close rate}: Percentage of users who switched off the notification toggle from being switched on within a day.
\end{itemize}

Our main business goals include duration and DAU, and the switch close rate needs to be neutral to ensure user satisfaction.

Results in \cref{tab:ab} shows that compared with uniform delivery, TIM increases the CTR of notifications by 1\%, resulting in a 0.1\% increase in DAU and a 0.066\% increase in watch time. At the same time, the volume of notifications (+0.023\% with p-value=0.48) and switch close rate (+0.448\% with p-value=0.47) remain neutral. This suggests that TIM guarantees both precise timing of notifications and business health, providing users with a strong sense of awareness.

\begin{table}[h]
    \caption{TIM's business gain in online A/B test, compared with uniform delivery. In parentheses are p-values.}
    \begin{tabular}{c|cc}
        \hline
        Metrics           & overall        & TIM            \\
        \hline
        DAU               & +0.026\%(0.37) & +0.100\%(0.00) \\
        send volume       & -0.214\%(0.00) & +0.023\%(0.48) \\
        CTR               & +0.437\%(0.00) & +1.217\%(0.00) \\
        watch time        & +0.014\%(0.73) & +0.066\%(0.11) \\
        switch close rate & +0.408\%(0.53) & +0.448\%(0.47) \\
        \hline
    \end{tabular}
    \label{tab:ab}
\end{table}
\section{Conclusion}
\label{sec:con}
In this paper, we introduce the Temporal Interaction Model (TIM). TIM uses historical data on user interactions over a long time, applies TAU to identify patterns in user behavior, and predicts CTR for each user in each time slot throughout the day. Furthermore, we present a strategy that optimizes the overall timing of notifications in mobile applications, leading to an enhancement of user engagement without causing excessive disturbance. Our experimental results demonstrate the effectiveness of TIM in enhancing notification delivery. In future work, we will explore the application of TIM in diverse scenarios and conduct targeted improvements, such as enhancing real-time performance and multi-objective estimation.

\bibliographystyle{ACM-Reference-Format}
\balance
\bibliography{sample-base}


\begin{thebibliography}{24}


\ifx \showCODEN    \undefined \def \showCODEN     #1{\unskip}     \fi
\ifx \showDOI      \undefined \def \showDOI       #1{#1}\fi
\ifx \showISBNx    \undefined \def \showISBNx     #1{\unskip}     \fi
\ifx \showISBNxiii \undefined \def \showISBNxiii  #1{\unskip}     \fi
\ifx \showISSN     \undefined \def \showISSN      #1{\unskip}     \fi
\ifx \showLCCN     \undefined \def \showLCCN      #1{\unskip}     \fi
\ifx \shownote     \undefined \def \shownote      #1{#1}          \fi
\ifx \showarticletitle \undefined \def \showarticletitle #1{#1}   \fi
\ifx \showURL      \undefined \def \showURL       {\relax}        \fi
\providecommand\bibfield[2]{#2}
\providecommand\bibinfo[2]{#2}
\providecommand\natexlab[1]{#1}
\providecommand\showeprint[2][]{arXiv:#2}

\bibitem[Bahdanau et~al\mbox{.}(2014)]%
        {bahdanau2014neural}
\bibfield{author}{\bibinfo{person}{Dzmitry Bahdanau}, \bibinfo{person}{Kyunghyun Cho}, {and} \bibinfo{person}{Yoshua Bengio}.} \bibinfo{year}{2014}\natexlab{}.
\newblock \showarticletitle{Neural machine translation by jointly learning to align and translate}.
\newblock \bibinfo{journal}{\emph{arXiv preprint arXiv:1409.0473}} (\bibinfo{year}{2014}).
\newblock


\bibitem[Chowdhury et~al\mbox{.}(2021)]%
        {chowdhury2021ceam}
\bibfield{author}{\bibinfo{person}{Farhan~Asif Chowdhury}, \bibinfo{person}{Yozen Liu}, \bibinfo{person}{Koustuv Saha}, \bibinfo{person}{Nicholas Vincent}, \bibinfo{person}{Leonardo Neves}, \bibinfo{person}{Neil Shah}, {and} \bibinfo{person}{Maarten~W Bos}.} \bibinfo{year}{2021}\natexlab{}.
\newblock \showarticletitle{CEAM: the effectiveness of cyclic and ephemeral attention models of user behavior on social platforms}. In \bibinfo{booktitle}{\emph{Proceedings of the international AAAI conference on web and social media}}, Vol.~\bibinfo{volume}{15}. \bibinfo{pages}{117--128}.
\newblock


\bibitem[De~Caigny et~al\mbox{.}(2018)]%
        {de2018new}
\bibfield{author}{\bibinfo{person}{Arno De~Caigny}, \bibinfo{person}{Kristof Coussement}, {and} \bibinfo{person}{Koen~W De~Bock}.} \bibinfo{year}{2018}\natexlab{}.
\newblock \showarticletitle{A new hybrid classification algorithm for customer churn prediction based on logistic regression and decision trees}.
\newblock \bibinfo{journal}{\emph{European Journal of Operational Research}} \bibinfo{volume}{269}, \bibinfo{number}{2} (\bibinfo{year}{2018}), \bibinfo{pages}{760--772}.
\newblock


\bibitem[Devlin et~al\mbox{.}(2018)]%
        {devlin2018bert}
\bibfield{author}{\bibinfo{person}{Jacob Devlin}, \bibinfo{person}{Ming-Wei Chang}, \bibinfo{person}{Kenton Lee}, {and} \bibinfo{person}{Kristina Toutanova}.} \bibinfo{year}{2018}\natexlab{}.
\newblock \showarticletitle{BERT: Pre-training of Deep Bidirectional Transformers for Language Understanding}.
\newblock \bibinfo{journal}{\emph{arXiv preprint arXiv:1810.04805}} (\bibinfo{year}{2018}).
\newblock


\bibitem[Dosovitskiy et~al\mbox{.}(2020)]%
        {dosovitskiy2020image}
\bibfield{author}{\bibinfo{person}{Alexey Dosovitskiy}, \bibinfo{person}{Lucas Beyer}, \bibinfo{person}{Alexander Kolesnikov}, \bibinfo{person}{Dirk Weissenborn}, \bibinfo{person}{Xiaohua Zhai}, \bibinfo{person}{Thomas Unterthiner}, \bibinfo{person}{Mostafa Dehghani}, \bibinfo{person}{Matthias Minderer}, \bibinfo{person}{Georg Heigold}, \bibinfo{person}{Sylvain Gelly}, {et~al\mbox{.}}} \bibinfo{year}{2020}\natexlab{}.
\newblock \showarticletitle{An Image is Worth 16x16 Words: Transformers for Image Recognition at Scale}. In \bibinfo{booktitle}{\emph{International Conference on Learning Representations}}.
\newblock


\bibitem[Gupta et~al\mbox{.}(2016)]%
        {gupta2016email}
\bibfield{author}{\bibinfo{person}{Rupesh Gupta}, \bibinfo{person}{Guanfeng Liang}, \bibinfo{person}{Hsiao-Ping Tseng}, \bibinfo{person}{Ravi~Kiran Holur~Vijay}, \bibinfo{person}{Xiaoyu Chen}, {and} \bibinfo{person}{R{\'o}mer Rosales}.} \bibinfo{year}{2016}\natexlab{}.
\newblock \showarticletitle{Email volume optimization at LinkedIn}. In \bibinfo{booktitle}{\emph{Proceedings of the 22nd ACM SIGKDD International Conference on Knowledge Discovery and Data Mining}}. \bibinfo{pages}{97--106}.
\newblock


\bibitem[Kamal et~al\mbox{.}(2021)]%
        {kamal2021hybrid}
\bibfield{author}{\bibinfo{person}{Rashid Kamal}, \bibinfo{person}{Paul McCullagh}, \bibinfo{person}{Ian Cleland}, {and} \bibinfo{person}{Chris Nugent}.} \bibinfo{year}{2021}\natexlab{}.
\newblock \showarticletitle{A hybrid model based on behavioural and situational context to detect best time to deliver notifications on mobile devices}.
\newblock In \bibinfo{booktitle}{\emph{Integrated Emerging Methods of Artificial Intelligence \& Cloud Computing}}. \bibinfo{publisher}{Springer}, \bibinfo{pages}{11--21}.
\newblock


\bibitem[Li et~al\mbox{.}(2023)]%
        {li2023digmn}
\bibfield{author}{\bibinfo{person}{Feifan Li}, \bibinfo{person}{Lun Du}, \bibinfo{person}{Qiang Fu}, \bibinfo{person}{Shi Han}, \bibinfo{person}{Yushu Du}, \bibinfo{person}{Guangming Lu}, {and} \bibinfo{person}{Zi Li}.} \bibinfo{year}{2023}\natexlab{}.
\newblock \showarticletitle{DIGMN: Dynamic Intent Guided Meta Network for Differentiated User Engagement Forecasting in Online Professional Social Platforms}. In \bibinfo{booktitle}{\emph{Proceedings of the Sixteenth ACM International Conference on Web Search and Data Mining}}. \bibinfo{pages}{384--392}.
\newblock


\bibitem[Liu et~al\mbox{.}(2019)]%
        {liu2019characterizing}
\bibfield{author}{\bibinfo{person}{Yozen Liu}, \bibinfo{person}{Xiaolin Shi}, \bibinfo{person}{Lucas Pierce}, {and} \bibinfo{person}{Xiang Ren}.} \bibinfo{year}{2019}\natexlab{}.
\newblock \showarticletitle{Characterizing and forecasting user engagement with in-app action graph: A case study of snapchat}. In \bibinfo{booktitle}{\emph{Proceedings of the 25th ACM SIGKDD International Conference on Knowledge Discovery \& Data Mining}}. \bibinfo{pages}{2023--2031}.
\newblock


\bibitem[Mehrotra et~al\mbox{.}(2015)]%
        {mehrotra2015designing}
\bibfield{author}{\bibinfo{person}{Abhinav Mehrotra}, \bibinfo{person}{Mirco Musolesi}, \bibinfo{person}{Robert Hendley}, {and} \bibinfo{person}{Veljko Pejovic}.} \bibinfo{year}{2015}\natexlab{}.
\newblock \showarticletitle{Designing content-driven intelligent notification mechanisms for mobile applications}. In \bibinfo{booktitle}{\emph{Proceedings of the 2015 ACM International Joint Conference on Pervasive and Ubiquitous Computing}}. \bibinfo{pages}{813--824}.
\newblock


\bibitem[O'Brien et~al\mbox{.}(2022)]%
        {o2022should}
\bibfield{author}{\bibinfo{person}{Conor O'Brien}, \bibinfo{person}{Huasen Wu}, \bibinfo{person}{Shaodan Zhai}, \bibinfo{person}{Dalin Guo}, \bibinfo{person}{Wenzhe Shi}, {and} \bibinfo{person}{Jonathan~J Hunt}.} \bibinfo{year}{2022}\natexlab{}.
\newblock \showarticletitle{Should i send this notification? Optimizing push notifications decision making by modeling the future}.
\newblock \bibinfo{journal}{\emph{arXiv preprint arXiv:2202.08812}} (\bibinfo{year}{2022}).
\newblock


\bibitem[Park et~al\mbox{.}(2017)]%
        {park2017don}
\bibfield{author}{\bibinfo{person}{Chunjong Park}, \bibinfo{person}{Junsung Lim}, \bibinfo{person}{Juho Kim}, \bibinfo{person}{Sung-Ju Lee}, {and} \bibinfo{person}{Dongman Lee}.} \bibinfo{year}{2017}\natexlab{}.
\newblock \showarticletitle{Don't bother me. I'm socializing! A breakpoint-based smartphone notification system}. In \bibinfo{booktitle}{\emph{Proceedings of the 2017 ACM Conference on Computer Supported Cooperative Work and Social Computing}}. \bibinfo{pages}{541--554}.
\newblock


\bibitem[Pielot et~al\mbox{.}(2017)]%
        {pielot2017beyond}
\bibfield{author}{\bibinfo{person}{Martin Pielot}, \bibinfo{person}{Bruno Cardoso}, \bibinfo{person}{Kleomenis Katevas}, \bibinfo{person}{Joan Serr{\`a}}, \bibinfo{person}{Aleksandar Matic}, {and} \bibinfo{person}{Nuria Oliver}.} \bibinfo{year}{2017}\natexlab{}.
\newblock \showarticletitle{Beyond interruptibility: Predicting opportune moments to engage mobile phone users}.
\newblock \bibinfo{journal}{\emph{Proceedings of the ACM on Interactive, Mobile, Wearable and Ubiquitous Technologies}} \bibinfo{volume}{1}, \bibinfo{number}{3} (\bibinfo{year}{2017}), \bibinfo{pages}{1--25}.
\newblock


\bibitem[Prabhakar et~al\mbox{.}(2022)]%
        {prabhakar2022multi}
\bibfield{author}{\bibinfo{person}{Prakruthi Prabhakar}, \bibinfo{person}{Yiping Yuan}, \bibinfo{person}{Guangyu Yang}, \bibinfo{person}{Wensheng Sun}, {and} \bibinfo{person}{Ajith Muralidharan}.} \bibinfo{year}{2022}\natexlab{}.
\newblock \showarticletitle{Multi-objective Optimization of Notifications Using Offline Reinforcement Learning}. In \bibinfo{booktitle}{\emph{Proceedings of the 28th ACM SIGKDD Conference on Knowledge Discovery and Data Mining}}. \bibinfo{pages}{3752--3760}.
\newblock


\bibitem[Su(2020)]%
        {linearatt}
\bibfield{author}{\bibinfo{person}{Jianlin Su}.} \bibinfo{year}{2020}\natexlab{}.
\newblock \bibinfo{booktitle}{\emph{Exploration of Linear Attention: Does Attention Need a Softmax?}}
\newblock
\urldef\tempurl%
\url{https://spaces.ac.cn/archives/7546}
\showURL{%
Retrieved Jan 30, 2024 from \tempurl}


\bibitem[Su et~al\mbox{.}(2024)]%
        {su2024roformer}
\bibfield{author}{\bibinfo{person}{Jianlin Su}, \bibinfo{person}{Murtadha Ahmed}, \bibinfo{person}{Yu Lu}, \bibinfo{person}{Shengfeng Pan}, \bibinfo{person}{Wen Bo}, {and} \bibinfo{person}{Yunfeng Liu}.} \bibinfo{year}{2024}\natexlab{}.
\newblock \showarticletitle{Roformer: Enhanced transformer with rotary position embedding}.
\newblock \bibinfo{journal}{\emph{Neurocomputing}}  \bibinfo{volume}{568} (\bibinfo{year}{2024}), \bibinfo{pages}{127063}.
\newblock


\bibitem[Tang et~al\mbox{.}(2020)]%
        {tang2020knowing}
\bibfield{author}{\bibinfo{person}{Xianfeng Tang}, \bibinfo{person}{Yozen Liu}, \bibinfo{person}{Neil Shah}, \bibinfo{person}{Xiaolin Shi}, \bibinfo{person}{Prasenjit Mitra}, {and} \bibinfo{person}{Suhang Wang}.} \bibinfo{year}{2020}\natexlab{}.
\newblock \showarticletitle{Knowing your fate: Friendship, action and temporal explanations for user engagement prediction on social apps}. In \bibinfo{booktitle}{\emph{Proceedings of the 26th ACM SIGKDD international conference on knowledge discovery \& data mining}}. \bibinfo{pages}{2269--2279}.
\newblock


\bibitem[Vaswani et~al\mbox{.}(2017)]%
        {vaswani2017attention}
\bibfield{author}{\bibinfo{person}{Ashish Vaswani}, \bibinfo{person}{Noam Shazeer}, \bibinfo{person}{Niki Parmar}, \bibinfo{person}{Jakob Uszkoreit}, \bibinfo{person}{Llion Jones}, \bibinfo{person}{Aidan~N Gomez}, \bibinfo{person}{{\L}ukasz Kaiser}, {and} \bibinfo{person}{Illia Polosukhin}.} \bibinfo{year}{2017}\natexlab{}.
\newblock \showarticletitle{Attention is all you need}. In \bibinfo{booktitle}{\emph{Proceedings of the 31st International Conference on Neural Information Processing Systems}}. \bibinfo{pages}{6000--6010}.
\newblock


\bibitem[Yang et~al\mbox{.}(2018)]%
        {yang2018know}
\bibfield{author}{\bibinfo{person}{Carl Yang}, \bibinfo{person}{Xiaolin Shi}, \bibinfo{person}{Luo Jie}, {and} \bibinfo{person}{Jiawei Han}.} \bibinfo{year}{2018}\natexlab{}.
\newblock \showarticletitle{I know you'll be back: Interpretable new user clustering and churn prediction on a mobile social application}. In \bibinfo{booktitle}{\emph{Proceedings of the 24th ACM SIGKDD International Conference on Knowledge Discovery \& Data Mining}}. \bibinfo{pages}{914--922}.
\newblock


\bibitem[Yang et~al\mbox{.}(2023)]%
        {yang2023deeppvisit}
\bibfield{author}{\bibinfo{person}{Guangyu Yang}, \bibinfo{person}{Efrem Ghebreab}, \bibinfo{person}{Jiaxi Xu}, \bibinfo{person}{Xianen Qiu}, \bibinfo{person}{Yiping Yuan}, {and} \bibinfo{person}{Wensheng Sun}.} \bibinfo{year}{2023}\natexlab{}.
\newblock \showarticletitle{DeepPvisit: A Deep Survival Model for Notification Management}. In \bibinfo{booktitle}{\emph{Companion Proceedings of the ACM Web Conference 2023}}. \bibinfo{pages}{973--977}.
\newblock


\bibitem[Yang et~al\mbox{.}(2017)]%
        {yang2017personalized}
\bibfield{author}{\bibinfo{person}{Lin Yang}, \bibinfo{person}{Mingxuan Yuan}, \bibinfo{person}{Yanjiao Chen}, \bibinfo{person}{Wei Wang}, \bibinfo{person}{Qian Zhang}, {and} \bibinfo{person}{Jia Zeng}.} \bibinfo{year}{2017}\natexlab{}.
\newblock \showarticletitle{Personalized user engagement modeling for mobile videos}.
\newblock \bibinfo{journal}{\emph{Computer Networks}}  \bibinfo{volume}{126} (\bibinfo{year}{2017}), \bibinfo{pages}{256--267}.
\newblock


\bibitem[Yuan et~al\mbox{.}(2022)]%
        {yuan2022offline}
\bibfield{author}{\bibinfo{person}{Yiping Yuan}, \bibinfo{person}{Ajith Muralidharan}, \bibinfo{person}{Preetam Nandy}, \bibinfo{person}{Miao Cheng}, {and} \bibinfo{person}{Prakruthi Prabhakar}.} \bibinfo{year}{2022}\natexlab{}.
\newblock \showarticletitle{Offline reinforcement learning for mobile notifications}. In \bibinfo{booktitle}{\emph{Proceedings of the 31st ACM International Conference on Information \& Knowledge Management}}. \bibinfo{pages}{3614--3623}.
\newblock


\bibitem[Yuan et~al\mbox{.}(2019)]%
        {yuan2019state}
\bibfield{author}{\bibinfo{person}{Yiping Yuan}, \bibinfo{person}{Jing Zhang}, \bibinfo{person}{Shaunak Chatterjee}, \bibinfo{person}{Shipeng Yu}, {and} \bibinfo{person}{Romer Rosales}.} \bibinfo{year}{2019}\natexlab{}.
\newblock \showarticletitle{A state transition model for mobile notifications via survival analysis}. In \bibinfo{booktitle}{\emph{Proceedings of the Twelfth ACM International Conference on Web Search and Data Mining}}. \bibinfo{pages}{123--131}.
\newblock


\bibitem[Zhang et~al\mbox{.}(2022)]%
        {zhang2022counterfactual}
\bibfield{author}{\bibinfo{person}{Guozhen Zhang}, \bibinfo{person}{Jinwei Zeng}, \bibinfo{person}{Zhengyue Zhao}, \bibinfo{person}{Depeng Jin}, {and} \bibinfo{person}{Yong Li}.} \bibinfo{year}{2022}\natexlab{}.
\newblock \showarticletitle{A Counterfactual modeling framework for churn prediction}. In \bibinfo{booktitle}{\emph{Proceedings of the Fifteenth ACM International Conference on Web Search and Data Mining}}. \bibinfo{pages}{1424--1432}.
\newblock


\end{thebibliography}

\end{document}